\documentclass[aps,onecolumn,nopacs,nofootinbib,floatfix,superscriptaddress]{revtex4}
\usepackage{graphicx}
\usepackage{amsfonts}
\usepackage{amssymb}
\usepackage{amsbsy}
\usepackage{hyperref}
\usepackage{amsmath}
\usepackage{mathrsfs}
\usepackage{latexsym}
\usepackage{natbib}
\usepackage{bm}
\usepackage{subfigure} 
\usepackage{color}
\usepackage{wasysym}
\usepackage{mathbbol}
\usepackage{bigints}
\allowdisplaybreaks
\usepackage[normalem]{ulem}
\usepackage[dvipsnames]{xcolor}
\usepackage{multirow}
\usepackage{physics}
\usepackage{comment}

\definecolor{napiergreen}{rgb}{0.16, 0.5, 0.0}

\begin{document}

\title{A note on entropy of matter in presence of gravity: status of extensivity of entropy}

\author{Saurav Samanta}
\email{srvsmnt@gmail.com}
\affiliation{Department of Physics, Bajkul Milani Mahavidyalaya, P. O. - Kismat Bajkul,
Dist. - Purba Medinipur, Pin - 721655, India}

\author{Bibhas Ranjan Majhi}
\email{bibhas.majhi@iitg.ac.in}
\affiliation{Department of Physics, Indian Institute of Technology Guwahati, Guwahati 781039, Assam, India}

\begin{abstract}
Entropy of matter in a very strong gravity depends on cross-sectional area of the container of the system -- is being further bolstered by calculating entropy of a monoatomic gas kept under uniform strong gravity at Newtonian scale. This bypasses the earlier analysis where existence of horizon is crucial. Also the extensivity of this entropy has been discussed in the light of the same of a two-space dimensional ultra-relativistic non-interacting gas without gravity. The whole analysis, as far as entropy is concerned, indicates that under strong gravity the microscopic degrees of freedom of the system are effectively contributed by the cross-sectional area of the system. 


\end{abstract}

\date{\today}

\maketitle

\section{Introduction}
Gravity and matter both influences each other -- equations of motions of metric tensor $g_{ab}$ (like Einstein's equations of motion) sourced by energy-momentum tensor $T_{ab}$ of matter determines how matter influences gravity, while field equations (like Klein-Gordon equation) shows that of gravity on matter. Although quantum theory of matter in presence of gravity is widely explored, but the quantum theory of (strong) gravity yet to be achieved. In this perspective, black holes are believed to play a crucial role in understanding the quantum nature of gravity. Ever since the idea of Bekenstein \cite{Bekenstein:1973ur} and Hawking \cite{Hawking:1975vcx} -- the black holes are known to behave as thermodynamic systems and the investigation is done in the direction of understanding the nature of underlying gravitational microscopic degrees of freedom. Although thermodynamics of horizon has been extensively studied, dependence of entropy on black hole horizon area (in Einstein's theory) isolates such thermodynamical system from the usual ones. Particularly the horizon entropy is not extensive in nature and apparently as a consequence the idea of Gibbs paradox is not well posed within this context. Understanding of the later one is very important as it revels whether the underlying microstructure is distinguishable or indistinguishable.  
However there have been two approaches based on microscopic counting of degrees of freedom of black hole -- one is loop quantum gravity (LQG) and other one is string theory (ST). It was observed that LQG approach is accustomed with the distinguishable nature of microstates while ST gives correct area dependence through indistinguishable counting (see the discussion in \cite{Kiefer:2007uw,Pithis:2012xw}). Moreover, a semiclassical approach \cite{Kiefer:2007uw} based on the work by Bekenstein and Mukhanov \cite{Bekenstein:1995ju} indicates the importance of Gibbs factor in incorporating the leading order correction (given by logarithmic of horizon area) in the black hole entropy. Another curious fact that the thermodynamics of the Hawking radiated particles is similar to that of usual $(1+1)$ dimensional matter, indicates the one dimensional behavior of quantum black holes \cite{Bekenstein:2001tj} (see also \cite{Nation:2010fd,Chen:2011mg,Mirza:2014vsa,Hod:2015dpa,Majhi:2015jua,Hod:2016xbu}).    

On the other hand the response of matter under gravity is equally interesting and important. Apart from the dynamics of matter influenced by gravity, its thermodynamics attracted a lot of attention.
In fact the idea of black hole thermodynamics is based on the quantization of matter on a classical black hole spacetime. Therefore we can assume that the knowledge of thermodynamic properties of matter can illuminate that of black hole. Hence a clear understanding of thermodynamics of external matter on a black hole background is also necessary and equally important. In this regard, an extensive study (mainly the entropy) has been done in literature \cite{Kolekar:2010py,Bhattacharya:2017bpl} (also see \cite{Kothawala:2011fm,Sourtzinou:2022xhd}). The most intriguing aspect of those studies is that the entropy of a box of gas shows distinct properties when it is Planck length $L_p$ far from the black hole horizon compared to the situation when the box is far away from the black hole. Particularly, the entropy depends on the volume of the box when the box is far away from the horizon, while the same depends on the surface area of the box facing the horizon when that surface is very close to a black hole. This near horizon feature of matter entropy has been tested for Rindler \cite{Kolekar:2010py}, event (static and stationary) \cite{Kolekar:2010py,Bhattacharya:2017bpl} as well as cosmological \cite{Bhattacharya:2017bpl} horizons and the dependence of entropy on area appears to be universal. Interestingly, this area dependence is similar to the feature shown by the entropy of black hole in Einstein's gravity. Therefore the microscopic degrees of freedom of the gas behaves distinctly in these two regions. Moreover, although the matter entropy for far away system is extensive in nature, the same in the near horizon may not be so. 

In this short note our aim is two-fold. We first like to inquire whether this area dependence of matter-entropy is due to the existence of horizon or it naturally arises under any generalized strong gravity set-up. Secondly, we want to find a concrete answers to the question -- whether the near horizon entropy of the box of gas is extensive in nature. In fact looking at the area dependence, in literature \cite{Kolekar:2010py} the entropy is declared as non-extensive. However we feel that just area dependence is not enough to make a conclusion. In particular we like to investigate the near horizon entropy of the box of gas through the Gibbs idea which, as far as we know, has not been done. Specifically we are curious to look for a possible condition which leads to the joint entropy of two boxes of gas as addition of individual entropies upon use of Gibbs factor. We already mentioned the importance of the use of Gibbs factor in understanding the nature of microscopic degrees of freedom -- whether they are distinguishable or indistinguishable. Therefore a study through the Gibbs idea is important to know whether the far away indistinguishable degrees of freedom remains same when the box arrives very near to the horizon.

Following the above motivations we first investigate the thermodynamics of a gas, contained in a cylinder. The gravity acting on each gas molecule is taken to be uniform in space given by Newtonian mechanics. We particularly concentrate on the entropy of the system. Interestingly, the entropy depends on the cross-sectional area of the cylinder for very strong gravitational acceleration. While for weak gravitational field the entropy follows usual volume dependency. As shown in \cite{Kolekar:2010py,Bhattacharya:2017bpl}, the area dependency was observed when the matter system was kept very near to the horizon. Therefore we feel that the area dependency is very much related to the strength of the gravity, rather than existence of a horizon. Hence the influence of strong gravity on matter thermodynamics is completely distinct. In this case the microscopic degrees of freedom of matter are effectively contributed from the cross-sectional area, even at the non-relativistic scale. 

Next we investigate the additivity property of entropy by considering a composite system consists of two subsystems of gas with volume $V_1$ and $V_2$, containing number of particles $N_1$ and $N_1$, respectively. These are in thermal equilibrium with the common temperature $T$. 
They are separated by a partition that allows only heat conduction, so that the subsystems are in thermal equilibrium with each other. The entropies of the individual subsystems are already known for two extreme conditions -- one when the system is in weak gravity and another when the system is in very strong gravity. Using these results, we find that introduction of the Gibbs factor $N!$ is sufficient to establish the additivity property of the entropy. In weak gravity situation, the total entropy is given by $S=S_1+S_1$ provided $N/V$ is same for the individual subsystems. This is expected as the entropy depends on volume of the system and hence conventional nature should follow. However the other case is non-trivial as the entropy depends on transverse area of the box. We find that for the strong gravity situation, the additivity property with the Gibbs factor follows provided $N/A$ is same for the individual subsystems. This is very interesting as it shows that the near-horizon entropy of matter could be extensive in nature.

In investigating the required condition $N/A = $ constant for extensiveness of matter entropy in the strong gravity regime we compare this with a $(2+1)$-dimensional thermodynamic system in Minkowski spacetime. It can be observed that the entropy of such a thermodynamics system is also extensive in nature provided $A/N$ does not change. Therefore as far as dependence of entropy on area is concerned, the matter system is effectively behaving like a $(2+1)$-dimensional system under strong gravity while its behavior is $(3+1)$-dimensional when the system is under weak gravity. Hence we conclude that the entropy of matter is always extensive (contrary to the previous claim in \cite{Kolekar:2010py}), however, the microscopic degrees of freedom are now living effectively in $(2+1)$-dimension rather than in $(3+1)$-dimension. In this respect another noticeable feature is that although the thermodynamics of Hawking radiated particle is effectively $(1+1)$-dimensional (hence the black hole is called as $(1+1)$-dimensional emitter), the effective behaviour of matter thermodynamics is one more space dimension as far as extensivity of entropy is concerned. However the dependence of entropy on temperature in strong gravity is different from usual $(2+1)$-dimensional case.

Let us now proceed to establish the claims.

\section{Canonical ensemble: a container of gas under gravity}
The classical canonical ensemble of thermodynamic system is described by the number of particles $N$, its volume $V$ and equilibrium inverse temperature $\beta$. To determine the other thermodynamic quantities one needs to concentrate on the partition function. The single-particle partition function is determined by \cite{Pathria:1996hda}
\begin{eqnarray}
Z_1(\beta)=\int e^{-\beta E}g(E)dE=\int e^{-\beta E} d(P(E))~,
\label{partition}
\end{eqnarray}  
where the relation between the density of states $g(E)$ and the phase space volume $P(E)$ is $g(E)=\frac{dP(E)}{dE}$. If all the particles are non-interacting (e.g. ideal gas), then the partition function for the whole system is given by 
\begin{eqnarray}
Z_N=\frac{1}{N!}[Z_1(\beta)]^N~.
\label{ZN}
\end{eqnarray}
In the above the factor $N!$ in the denominator is introduced by considering that the particles are indistinguishable. Its importance will be visible latter. This has been incorporated by Gibbs to make the entropy of the system extensive in nature and therefore is called as the Gibbs factor. 
Using Stirling approximation $\ln N!\simeq N\ln N-N$ one obtains
$\ln Z_N=N\ln \frac{Z_1}{N}+N$. Then the average energy of the system
 is determined as 
 \begin{eqnarray}
\bar{E}=-\frac{\partial \ln Z_N}{\partial \beta}=-\frac{N}{Z_1}\frac{\partial Z_1}{\partial \beta}~. 
\label{E}
\end{eqnarray}
The entropy of the system is then given as
\begin{eqnarray}
S=\ln Z_N+\beta \bar{E}=N\ln \frac{Z_1}{N}+N+\beta \bar{E}~.
\label{S}
\end{eqnarray} 
The pressure of the system is defined as
\begin{eqnarray}
P=\frac{1}{\beta}\frac{\partial S}{\partial V}~.
\end{eqnarray}

Recall that our main attention is to investigate the effect of gravity on the entropy of a thermodynamic system. We consider (3+1) spacetime dimensions and take Newtonian gravity for simplicity. We will see that such a simple model can lead to various interesting consequences depending upon the strength of gravity and this will illuminate one of our main goals. We consider a monoatomic gas contained in a cylinder of base area $A$ and length $L$ placed in a uniform gravitational field (constant gravitational acceleration) with its axis parallel to the field. The box contains $N$ identical monoatomic gas molecules with equilibrium inverse temperature $\beta$. The total energy of a single particle is
\begin{eqnarray}
E=\frac{1}{2m}(p_x^2+p_y^2+p_z^2)+mgz=\frac{p^2}{2m}+mgz~.
\end{eqnarray}
$p_i$ and $m$ are the components of linear momentum and mass of each particle. $g$ is the gravitational acceleration and $z$ is the perpendicular distance of the particle from the base of the cylinder.
Then the single-particle partition function is
\begin{eqnarray}
Z_1=\int e^{-\beta(\frac{p^2}{2m}+mgz)}d^3qd^3p=\frac{(2\pi)^{3/2}m^{1/2}A}{g\beta^{5/2}}\left(1-e^{-\beta mgL}\right)~.
\label{B1}
\end{eqnarray}
The above result has already been calculated in \cite{Note1}. 

Before proceeding further, let us make some observations.
In (\ref{B1}) if we take the limit $g\rightarrow 0$ (no gravity), then the single-particle partition function reduces to
\begin{eqnarray}
Z_1\simeq\left(\frac{2\pi m}{\beta}\right)^{3/2}V~,
\end{eqnarray}
where $V=AL$ is the volume of the box. This result is well known in statistical mechanics as the single-particle partition function for free particle (i.e. without gravity). On the other hand for large value of $g$ (i.e. strong gravity) (\ref{B1}) takes the following form:
\begin{eqnarray}
Z_1\simeq\frac{(2\pi)^{3/2}m^{1/2}A}{g\beta^{5/2}}~.
\end{eqnarray}
Clearly in one case (zero gravity) $Z_1$ is proportional to volume and in other case (strong gravity) it is proportional to area. Also in these two cases, dependence on temperature is different. In the following analysis we show similar behavior for the system entropy.

Now using (\ref{E}) the average internal energy of the system is obtained as
\begin{eqnarray}
\bar{E}=\frac{5N}{2\beta}-\frac{mgNL}{e^{\beta mgL}-1}~.
\end{eqnarray}
Then using (\ref{S}) we find the entropy of the system:
\begin{eqnarray}
S=N\ln\left[\frac{(2\pi)^{3/2}m^{1/2}A}{gN\beta^{5/2}}\left(1-e^{-\beta mgL}\right)\right]+\frac{7N}{2}-\frac{mgNL\beta}{e^{\beta mgL}-1}~.\label{genentropy}
\end{eqnarray}
For $g\rightarrow 0$ limit, the above expression yields
\begin{eqnarray}
S\simeq N\ln\left[\frac{V}{N}\left(\frac{2\pi m}{\beta}\right)^{3/2}\right]+\frac{5N}{2}~.\label{Szero}
\end{eqnarray}
Once again this expression of entropy for ideal gas is well known in statistical mechanics.
On the other hand if the gravitational field is strong (i.e. $g$ is very large) then one finds
\begin{eqnarray}
S\simeq N\ln\left[\frac{(2\pi)^{3/2}m^{1/2}A}{gN\beta^{5/2}}\right]+\frac{7N}{2}~.
\label{B3}
\end{eqnarray}
The later entropy is a curious case. The strong gravity (even the Newtonian one) in this case makes the entropy of the matter dependent on cross-sectional area of the container, rather than its volume. Interestingly such area dependence has already been observed for a strong gravitational object, like black holes in Einstein's gravity.

Just for curiosity, we calculate the transverse and longitudinal pressures of the above system. These are obtained from
\begin{eqnarray}
P_{\perp}=\frac{1}{\beta L}\frac{\partial S}{\partial A}~; \,\,\,\,\,\   P_{\parallel}=\frac{1}{\beta A}\frac{\partial S}{\partial L}~.
\end{eqnarray}
Using the expression of entropy (\ref{genentropy}) we find
\begin{eqnarray}
P_{\perp}=\frac{N}{\beta V}~; \,\,\,\,\,\  P_{\parallel}=\frac{N\beta m^2g^2L}{A}\frac{e^{\beta mgL}}{\left(e^{\beta mgL}-1\right)^2}~.
\end{eqnarray}
Clearly these two pressures are different. In the zero gravity limit, $P_{\parallel}$ goes to $P_{\perp}$ and so in that case it is not necessary to distinguish between longitudinal and transverse pressures. This common pressure can easily be obtained from the expression of entropy (\ref{Szero}).
However in strong gravitational field, though $P_{\perp}$ remains same, $P_{\parallel}$ is actually zero.


Let us now shift our attention to the relativistic gravitational regime. In this case we consider a box of ultra-relativistic gas placed in a black hole spacetime. The entropy of the system has already been investigated in \cite{Kolekar:2010py,Bhattacharya:2017bpl}. It has been observed that the entropy of the gas, when the box is kept far from the horizon, depends on volume of the box. However, if the surface of the box, facing the horizon, is at Planck length $L_p$ distance away from the horizon, then the entropy curiously depends on the surface area of the box. This also holds for Rindler frame as well, which is a relativistic extension of uniform Newtonian gravity, adopted in the previous example. The explicit form of the near-horizon entropy in $(3+1)$-dimensional spacetime is given by \cite{Kolekar:2010py,Bhattacharya:2017bpl}{\footnote{The last factor in \cite{Kolekar:2010py,Bhattacharya:2017bpl} is obtained as $3$, instead of $4$. However if one incorporates the Gibbs factor in calculating the partition function of the system, then the correct factor must be $4$ (see Appendix \ref{App1} for the clarification). Here we consider the Gibbs factor for later purpose.}}
\begin{eqnarray}
S=N\left[\ln\left(\frac{4\pi L_p}{\beta_{\textrm{loc}}^3}\frac{A}{N}\right)+4\right]~,
\label{B2}
\end{eqnarray}
where $\beta_{\textrm{loc}}$ is the redshifted local (Tolman) temperature of the ultra-relativistic gas. It has been observed that the same is true for any static spherically symmetric or stationary black hole spacetime. Also, it holds for cosmological horizon as well.  

The interesting finding from these two situations (one with Newtonian gravity and another with relativistic gravity) is that in both cases the entropy of the considered matter depends on area when the gravity is very strong. For the first case, gravitational acceleration $g$ is very large and in other case, the thermodynamic system is near the horizon where the gravity is very strong. So it seems that the strength of gravity plays the key role in determining dependence of entropy on geometric dimension of the system. However, dependence on temperature varies case by case. 

\section{Thermodynamics of ultra-relativistic gas in 2-dimension}
To gain greater understanding about the properties of entropy of the system under strong gravity, we will investigate the status of extensive nature of this entropy. The appearance of area dependence questions the validity of the additive property of the same. In fact, in literature expression (\ref{B2}) is considered to be non-extensive. 
However, we feel more investigation is needed to make it conclusive. Since in this case the entropy is area dependent, we like to discuss how the entropy of a $(2+1)$-dimensional ultra-relativistic gas behaves without gravity and compare it with our $(3+1)$-dimensional strong gravitational situation, analyzed above. Therefore below we calculate the entropy of such a system.  

Consider a system of $N$ number of non-interacting ultra-relativistic particles ($E=p$), confined on a surface of area $A$. Here the surface is lying on the surface of the earth and hence no gravity is acting. Like earlier, the single-particle partition function is given by
\begin{eqnarray}
Z_1=\int e^{-\beta E}d^2qd^2p=\int d^2q\int e^{-\beta p}d^2p=\int d^2q\int e^{-\beta p}2\pi pdp~.
\end{eqnarray}
Integration of the space part is area $A$ of the system. Integration of the momentum part can be evaluated by gamma function. So we find
\begin{eqnarray}
Z_1=\frac{2\pi A}{\beta^2}~.
\end{eqnarray}
Then use of (\ref{S}) yields the entropy of the system as
\begin{eqnarray}
S=\ln Q_N+\beta E=N\ln\left(\frac{2\pi A}{\beta^2}\right)-N\ln N+3N=\ln \left(\frac{2\pi}{\beta^2}\frac{A}{N}\right)^N+3N 
\label{S1}
\end{eqnarray}
Here the entropy, as expected, depends on the area of the system. In this sense, the above one is similar to (\ref{B3}) and (\ref{B2}). However the temperature dependencies in all the expressions are totally different.

Now we consider two systems having $N_1$ and $N_2$ particles and same temperature but separated by a common boundary. The areas of the systems are $A_1$ and $A_2$, respectively. The total entropy of the system will be
\begin{eqnarray}
S_{\textrm{total}}=S_1+S_2=\ln \left(\frac{2\pi}{\beta^2}\frac{A_1}{N_1}\right)^{N_1}+3N_1+\ln \left(\frac{2\pi}{\beta^2}\frac{A_2}{N_2}\right)^{N_2}+3N_2~.
\end{eqnarray}
If the number density of particles (i.e. number of particles per unit area) in two systems are same $\left(i.e. ~\frac{N_1}{A_1}=\frac{N_2}{A_2}\right)$, above expression can be written as
\begin{eqnarray}
S_{\textrm{total}}=\ln \left(\frac{2\pi}{\beta^2}\frac{A_1+A_2}{N_1+N_2}\right)^{N_1+N_2}+3(N_1+N_2)~.
\end{eqnarray}
This implies that if we remove the common boundary of two systems, there will not be any change of entropy. Since both systems have same temperature and same (surface) density, after removal of the boundary there will not be any heat flow or diffusion. So mixing of particles does not constitute any physical process. This allows us to mentally divide a system into smaller subsystems. 

We observed that under strong gravity, the area dependence of entropies (\ref{B3}) and (\ref{B2}) is similar to (\ref{S1}). Hence the above discussion implies that (\ref{B3}) and (\ref{B2}) will be extensive in nature provided one imposes $N_1/A_1 = N_2/A_2$, instead of $N_1/V_1 = N_2/V_2$. The physical implication is as follows. Under very strong gravity, the microscopic degrees of freedom, liable to entropy of the system, are effectively contributed from the surface rather than from the volume. The reason can be put forwarded from the relativistic idea of length contraction. Due to gravity the longitudinal length along the direction of gravity is contracted and the contraction is more when the gravity is strong. Therefore, the system effectively behaves like one dimension less and holographic with respect to entropy. Therefore extensivity of the entropy of matter under strong gravity is much feasible under the condition $N/A$ is constant. Also note that in all the calculations the Gibbs factor $N!$ plays an important role; without this we do not have the extensivity of entropy. Therefore, in these cases, the particles under strong gravity remain indistinguishable as they are in zero gravity.

\section{Conclusion}

Strong gravity influences entropy of a matter system quite dramatically. In this case, entropy depends on the cross-sectional area of the container of the system. This was initially established through studying a box of gas near the black hole horizon. Here we consider a prototype model when a cylinder of gas is under Newtonian gravity where the gravitational acceleration is uniform. We found that the entropy of such system depends on the cross-sectional area of the cylinder under high gravitational acceleration. However, at the zero gravity limit the entropy depends on volume of the system. It provides more robust evidence in favor of area dependence of matter entropy under strong gravity as in this case there is no existence of horizon. 

To analyze this area dependence, we further computed the entropy of a $(2+1)$ non-interacting ultra-relativistic gas system in absence of gravity. In this case introduction of Gibbs factor is sufficient to establish the extensive property of entropy, provided $N/A$ remains same in all sub-systems. We observed exactly the same for the entropy under strong gravity as well. In this sense, the matter entropy in presence of strong gravity also satisfies extensivity property. Also such entropy behaves effectively like a $(2+1)$-dimensional system in absence of gravity. However, the dependence of entropy on temperature is different from $(2+1)$-dimensional case. It implies that under very strong gravity the microscopic degrees
of freedom, liable to entropy of the system, are effectively contributed from the surface rather than from the volume.

The above observations are based on few idealized models and therefore lacks generality. So to get full understanding of the underlying physics further investigations are required. However we hope the present discussion put a tiny light to this area and hence may be important for future progress.

\begin{appendix}
\section*{Appendix}
\section{Entropy of gas near the horizon}\label{App1} 
The single particle partition function for the box of ultra-relativistic gas, kept very near the horizon of a $(D+1)$ dimensional static spherically symmetric black hole, is calculated in \cite{Kolekar:2010py}. In $D$-space dimensions it is given by
\begin{equation}
Z_1 = \frac{D! \pi^{D/2}}{\Gamma(\frac{D}{2}+1)} \frac{A_{D-1}\Big(L_P/(D-1)\Big)}{(\beta_{\text{loc}})^D}~.
\label{A1}    
\end{equation}
In the above $A_{D-1}$ is the cross-sectional area.
Then including the Gibbs factor, the partition  function $Z_N$ for $N$ number of particles is calculated by Eq. (\ref{ZN}). The average energy is given by
\begin{equation}
\bar{E} = - \frac{\partial \ln Z_N}{\partial\beta_{\text{loc}}} = \frac{ND}{\beta_{\text{loc}}}~.  
\label{A2}
\end{equation}
So the entropy of the system is evaluated from (\ref{S}) as
\begin{eqnarray}
S &=& N\ln \frac{Z_1}{N} + N + \beta_{\text{loc}}\bar{E}
\nonumber\\
&=& N \ln\Big[\frac{D! \pi^{D/2}}{\Gamma(\frac{D}{2}+1)} \frac{A_{D-1}(L_P/(D-1))}{N(\beta_{\text{loc}})^D} \Big] + N(D+1)~.
\label{A3}    
\end{eqnarray}
For $D=3$, the above reduces to (\ref{B2}) with the identification $A_{2} = A$. Note that apart from a factor $N$ in the denominator of the first term, the appearance of the second term (the factor $N$) in the first equality is solely due to the inclusion of Gibbs factor. This provides an extra term in the last equality. Therefore the last terms is $N(D+1)$, instead of $ND$. Similarly for the other horizons the same argument follows. 
\end{appendix}

\bibliographystyle{utphys1.bst}

\bibliography{bibtexfile}

\end{document}